\documentclass[prd,twocolumn,a4paper,superscriptaddress,floatfix]{revtex4}
\usepackage{graphicx}
\begin{document}
\newcommand{\be}{\begin{equation}}
\newcommand{\ee}{\end{equation}}
\newcommand{\bq}{\begin{eqnarray}}
\newcommand{\eq}{\end{eqnarray}}
\newcommand{\bsq}{\begin{subequations}}
\newcommand{\esq}{\end{subequations}}
\newcommand{\bc}{\begin{center}}
\newcommand{\ec}{\end{center}}
\newcommand\lapp{\mathrel{\rlap{\lower4pt\hbox{\hskip1pt$\sim$}} \raise1pt\hbox{$<$}}}
\newcommand\gapp{\mathrel{\rlap{\lower4pt\hbox{\hskip1pt$\sim$}} \raise1pt\hbox{$>$}}}

\title{Dynamics of Biased Domain Walls and the Devaluation Mechanism}
\author{P. P. Avelino}
\email[Electronic address: ]{ppavelin@fc.up.pt}
\affiliation{Centro de F\'{\i}sica do Porto, Rua do Campo Alegre 687, 4169-007 P
orto, Portugal}
\affiliation{Departamento de F\'{\i}sica da Faculdade de Ci\^encias
da Universidade do Porto, Rua do Campo Alegre 687, 4169-007 Porto, Portugal}
\author{C.J.A.P. Martins}
\email[Electronic address: ]{Carlos.Martins@astro.up.pt}
\affiliation{Centro de Astrof\'{\i}sica, Universidade do Porto, Rua das Estrelas
 s/n, 4150-762 Porto, Portugal}
\affiliation{DAMTP, University of Cambridge, Wilberforce Road, Cambridge CB3 0WA
, United Kingdom}
\author{L. Sousa}
\email[Electronic address: ]{laragsousa@gmail.com}
\affiliation{Centro de F\'{\i}sica do Porto, Rua do Campo Alegre 687, 4169-007 P
orto, Portugal}
\affiliation{Departamento de F\'{\i}sica da Faculdade de Ci\^encias
da Universidade do Porto, Rua do Campo Alegre 687, 4169-007 Porto, Portugal}

\date{26 May 2008}
\begin{abstract}
We study the evolution of biased domain walls in the early universe. We explicitly discuss the roles played by the surface tension and volume pressure in the evolution of the walls, and quantify their effects by looking at the collapse of spherical wall solutions. We then apply our results to a particular mechanism, known as the devaluation scenario, in which the dynamics of biased domain walls was suggested as a possible solution to the cosmological constant problem. Our results indicate that devaluation will in general lead to values of the cosmological constant that differ by several orders of magnitude from the observationally inferred value, $\rho^{1/4}_{vac}\sim10^{-3} \, \rm eV$. We also argue that the reasons behind this are not specific to a particular realization, and are expected to persist in any scenario of this kind, except if a low energy cut-off on the spectra of vacuum energy densities, of the order of the critical density at the present time, is postulated. This implies that any such scenario will require a fine-tuning similar to the usual one.
\end{abstract}
\pacs{98.80.Cq, 11.27.+d, 98.80.Es}
\maketitle

\section{\label{intr}Introduction}

Topological defects necessarily form at cosmological phase transitions \cite{KIBBLE,VSH}. Work on defects tends to focus on cosmic strings, which are usually benign in terms of their cosmological consequences. Domain walls are comparatively neglected since they are subject to tight constraints, in particular the so called Zel'dovich bound \cite{ZEL}. Nevertheless, they can play interesting cosmological roles. For example, it was thought that a particular type of domain wall network that would be frozen by the cosmological expansion could be a candidate for the dark energy that is thought to be accelerating the recent universe \cite{SOLID}, although detailed studies have now ruled out this scenario \cite{IDEAL1,IDEAL2}. Another example is that of biased domain wall networks \cite{Gelmini,Larsson}, which may provide a way to evade the Zel'dovich bound and are also behind the recently proposed devaluation scenario \cite{Freese}.

In order to analyse this and other paradigms one needs to understand the dynamics of domain wall networks in some detail. Since they are intrinsically non-linear objects this usually requires a careful combination of analytic modeling and high-performance numerical simulation. On the other hand, this toolkit can be applied not only to the simplest wall networks \cite{SIMS1,SIMS2,AWALL} but also to more complicated ones, such as walls with junctions \cite{IDEAL1,IDEAL2,IDEAL3}. In the present report we will take a detailed look at the dynamics of biased domain wall networks, and then use our results to carry out a thorough analysis of the devaluation scenario. Another recent analysis of this type of walls can be found in \cite{Lalak}. Throughout this paper we shall use natural units with $c=\hbar=k_B=1$.

\section{\label{dwevol} Basic domain wall evolution}

The space-time history of an infinitely thin domain wall can be represented as
\be
x^{\mu}=x^{\mu}(\zeta^a)\,, \qquad \mu=0,1,2,3\,, \qquad a=0,1,2\,.
\ee
The action describing the dynamics of such domain walls is given by
\be
S=-\sigma \int d^2 \zeta {\sqrt \beta}\,,
\ee
where $\sigma$ is the domain wall tension, $\beta=det(\beta_{ab})$ and $\beta_{ab}=g_{\mu \nu} x^\mu_{,a} x^\nu_{,b}$ is the world-sheet metric. The equation of motion of the domain walls is
\be
\label{eqmotion}
\beta^{-1/2} \partial_a({\sqrt \beta} \beta^{ab} x^\mu_{,b})+
\Gamma^\mu_{\nu \alpha}  \beta^{ab}  x^\nu_{,a}  x^\alpha_{,b}=0\,.
\ee
Although this is not easy to solve in general, we may simplify the problem considerably by considering the evolution of a spherical domain wall in a flat FRW universe with line element
\be
\label{3}
ds^2=d t^2 - a^2(t){\bf d x}^2\,,
\ee
where $a$ is the scale factor, $t$ is the physical time and ${\bf x}$ are conformal spatial coordinates. The space-time trajectory of a spherically symmetric domain wall can be written as
\be\label{4}
{\bf x}(t,\theta) = q(t) \,
\left(\cos \theta \cos \phi, \cos\theta \sin \phi, \sin \theta \right).
\ee
with $0\leq \phi \leq 2 \pi$, $0\leq \theta\leq \pi$ independent of $t$. Spherical domain walls in Minkowski space-time ($a=1$) have a conserved wall invariant area \cite{Loops} given by $S=4\pi \gamma q^2$ (with the usual definition of the Lorentz factor, $\gamma = (1-v^2)^{-1/2}$ with $v=a dq/dt$) which implies that $R = \gamma^{1/2} |q|$ is a conserved wall invariant radius (this is analogous to $R_s=\gamma |q|$ for a cosmic string loop). It is then straightforward to show that the spherical domain wall equation of motion in Minkowski space is given by
\be\label{1}
\frac{dv}{dt}=(1-v^2)\left(-\frac{2}{q}\right)\,.
\ee

For the more generic case with Hubble expansion the wall invariant radius includes the scale factor, $R = \gamma^{1/2} a q$, and the effect of the Hubble damping term follows trivially from the fact that the domain wall momentum per unit comoving area for an infinite planar domain wall is proportional to $a^{-1}$ (so that $v \gamma \propto a^{-3}$), leading to
\be
\frac{dv}{dt}=(1-v^2)\left(\frac{f(v)}{R}-3Hv\right)\,,
\label{8}
\ee
where, $H=(da/dt)/a$ is the Hubble parameter and we have defined $f(v)=2 \gamma^{1/2} sign(-q)$. The other effect of expansion is that $R$ is no longer a constant but becomes a dynamical quantity. It is also straightforward to show, for example using Eqn. (\ref{1}), that its evolution equation has the form
\be
\frac{dR}{dt}=\left(1-\frac{3}{2} v^2\right)HR\,.
\label{7}
\ee

It is instructive to compare these with the (averaged) evolution equations for a whole domain wall network \cite{AWALL}. These are written in terms of a characteristic length scale defined as
\be
\rho\equiv\frac{\sigma}{L}
\ee
and a root-mean squared velocity $v$, and have the form
\be
\frac{dL}{dt}=(1+3v^2)HL\label{vosl}
\ee
\be
\frac{dv}{dt}=(1-v^2)\left(\frac{k}{L}-3Hv\right)\,.\label{vosv}
\ee
For the moment we are ignoring the network's energy losses. The matching is closer than might naively appear if one recalls that in the context of a simple one-scale model the correlation length is the only length scale in the problem and is therefore identified with the curvature radius \cite{QUANT}, so even though the velocities in both cases are defined differently their evolution is the same except for the behavior of the phenomenological curvature parameter $k$ (which in the context of analytic modeling has a non-trivial velocity dependence \cite{AWALL,QUANT}).

On the other hand, when contrasting the length scales the key difference is that $L$ is to be understood as averaged over a network (hence a sizable volume) whereas $R$ in our above discussion essentially applies to a single defect. Let us consider a domain wall that separates domains $1$ (inside) and $2$ (outside), and assume that the volume fraction occupied by each vacuum is $f_1$ and $f_2$ with $f_1 \le f_2$. Then a definition of characteristic scale which is more directly related to the typical size of the domain walls is given by
\be
L_{\rm eff} \sim L \frac{f_1}{f_2}\,.
\label{leff}
\ee
In the limit $f_1 \ll f_2$ we expect to have only isolated domains so that $f_1/f_2 \propto L_{\rm eff}^3/a^3$ from which one obtains
\be
L_{\rm eff} \propto \frac{a^{3/2}}{L^{1/2}}\,. 
\ee
Because of this difference it is not possible to find a simple expression that smoothly interpolates between the two limits and is valid throughout: a wall network in such an intermediate regime will have highly non-trivial statistical properties. (Note that this is not a serious problem at all: the dynamics of biased walls is sufficiently fast to ensure that no network will ever be in such an intermediate state for a significant amount of time.) Substituting the above relation into Eqn. \ref{vosl} we get the suggestive
\be
\frac{dL_{\rm eff}}{dt}=\left(1-\frac{3}{2}v^2\right)HL_{\rm eff}\,.\label{voseff}
\ee
All the above discussion in essentially classical, in particular by neglecting energy losses through radiative mechanisms. These can still be added to the model phenomenologically when needed. However, for our present purposes we shall be more interested in modeling the effects of biasing, as described below.

\section{\label{bias}Biased domain walls}

In the simplest model of biased domain walls there is an asymmetry between the two minima of the potential \cite{Gelmini,Larsson}. The volume pressure from the biasing provides a further mechanism which affects the dynamics of these walls. Depending on its importance relative to other processes---most notably the surface tension---the walls may be long-lived (as in the standard case) or disappear almost immediately.

\subsection{Qualitative analysis}

Let us consider a simple model
\begin{equation}
{\cal L}=\frac{1}{2}(\partial_\mu\phi)(\partial^\mu\phi)-V(\phi)
\end{equation}
with the tilted potential
\begin{equation}
V(\phi)=\frac{\lambda}{4}\eta^4\left[\left(\frac{\phi^2}{\eta^2}-1\right)^2+\kappa\frac{\phi}{\eta}\right]\,.
\end{equation}
It then follows \cite{VSH} that the height of the potential barrier and surface tension are respectively
\begin{equation}
V_0=\frac{\lambda}{4}\eta^4\,
\end{equation}
\begin{equation}
\sigma\sim\sqrt{\lambda}\eta^3\,,
\end{equation}
while the wall thickness and the asymmetry parameter (or energy difference between the two vacua) are respectively
\begin{equation}
\delta\sim\frac{\eta}{\sqrt{V_0}}\sim(\sqrt{\lambda}\eta)^{-1}
\end{equation}
\begin{equation}
\epsilon=2\kappa V_0\,.
\end{equation}

Now, the surface pressure (from the tension force) and the volume pressure (from the energy difference between vacua) are
\begin{equation}
p_T=\frac{\sigma}{R}
\end{equation}
\begin{equation}
p_V=\epsilon\,.
\end{equation}
At early times the surface tension tends to dominate (due to the small curvature radii), but when the domains become large enough they will decay. This happens when \cite{Larsson}
\begin{equation}
R\sim\frac{\sigma}{\epsilon}\,.\label{decaybound}
\end{equation}
For example, we phenomenologically expect that at the Ginzburg temperature (when the network first becomes well-defined) its correlation length should approximately be given by $L\sim(\lambda\eta)^{-1}$ and in this case we get
\begin{equation}
\left(\frac{p_V}{p_T}\right)_{T_G}\sim\frac{\kappa}{2\sqrt{\lambda}}\,;
\end{equation}
typically one might expect this to be smaller but not much less than unity, but in principle there is enough parameter freedom to make it much larger or much smaller. When the volume pressure dominates, one expects that the walls will move with an acceleration
\begin{equation}
\frac{\epsilon}{\sigma}\sim\lambda^{1/2}\kappa\eta\,.\label{accel}
\end{equation}
On the other hand, if the surface pressure dominates initially the walls may survive long enough to reach a linear scaling regime, $L\sim t$. How fast the volume pressure becomes important, and hence how fast the walls can decay is a key issue in several cosmological scenarios, including devaluation. Moreover, the Zel'dovich bound \cite{ZEL} provides an additional (and often limiting) observational constraint.

\subsection{Analytic modeling}

Modeling the effect of this bias is fairly straightforward in the context of the analytic model we introduced above. Suppose that the two vacua on either side of the domain wall have different energy densities $V_{\rm in}$ and $V_{\rm out}$ and define $\epsilon = V_{\rm in}-V_{\rm out}$. The resulting effect on domain wall dynamics is very easy to understand if we consider a planar domain wall in Minkowski space. In this case the inner and outer regions of the domain wall are not well defined but we shall assume that the domain wall is moving in the direction ${\rm out} \to {\rm in}$. Energy conservation implies that
\be
d(\sigma \gamma) = v dt \epsilon\,,
\label{14}
\ee
from which we get
\be
\frac{dv}{dt} = \frac{\epsilon}{\sigma \gamma^3}\,;
\label{15}
\ee
we can immediately note that this is coincides with Eqn. (\ref{accel}) apart from the relativistic gamma factors. With our conventions, if $\epsilon > 0$ then the wall gains momentum. We thus confirm the expectation that the domain wall will feel a pressure which will tend to drive it into the region of higher energy density. We can alternatively write this as
\be
\frac{dv}{dt}=(1-v^2)^{3/2}\frac{\kappa\lambda^{1/2}}{2}\eta\equiv\frac{(1-v^2)^{3/2}}{R_v}\label{curvv}
\ee
which makes it clear that $\epsilon$ acts like an effective curvature, which therefore accelerates the wall. Its distinguishing feature is that the scale is set by the micro-physics of the model in question (ultimately by the form of the potential), rather than the macroscopic dynamics as is the case for the usual curvature radius of the walls. Moreover, this length scale is constant, whereas the usual curvature radius increases as the walls evolve. This implies that the volume pressure term gradually becomes more important. Having said that, note that the extra $\gamma$ factor can switch this term off if the walls become ultra-relativistic ($v\to1$). This discussion makes it clear that we can now add this volume pressure correction to our domain wall evolution equation (\ref{8}), yielding
\be
\frac{dv}{dt}=(1-v^2)\left(\frac{f(v)}{R} +\frac{\epsilon}{\sigma \gamma}-3Hv\right)\,,
\label{13}
\ee
or equivalently
\be
\frac{dv}{dt}=(1-v^2)\left(\frac{f(v)}{R} +\frac{\gamma^{-1}}{R_v}-3Hv\right)\,.
\label{13alt}
\ee

Taking into account the effect of the bias in Eqn. (\ref{8}), which describes the evolution of the wall invariant radius, $R$, we obtain
\be
\frac{dR}{dt}=\left(1-\frac{3}{2} v^2\right)HR+\frac{\epsilon vR}{2\sigma\gamma}\,,
\label{12}
\ee
or again using the definition of $R_v$ provided in Eqn. \ref{curvv} it can be written in the more suggestive form
\be
\frac{dR}{dt}=\left(1-\frac{3}{2} v^2\right)HR+(1-v^2)^{1/2}\frac{v}{2}\frac{R}{R_v}\,,
\label{12phys}
\ee
which should be compared with Eqn. (\ref{voseff}). Notice that this additional term due to the bias is associated with an extra energy loss by the network. As discussed above, the importance of this term is expected to grow as the wall evolves, although it switches off in the ultra-relativistic limit. 

It obviously follows from the above discussion that this model reproduces the result \cite{Larsson} that domains with a larger energy density will decay when their typical size $R\gapp\sigma/\epsilon$, and indeed it provides a more quantitative estimate---note that the relativistic gamma factor may be significant. We may also compare the importance of the pressure term with that of the Hubble damping term in Eqn. (\ref{8}). For non-relativistic domain walls the pressure term dominates over the Hubble damping term slightly earlier than over the curvature term (assuming that $R \lapp H^{-1}$). However, for $R \sim H^{-1}$ and $v$ not too small the two criteria are very similar. If $R \lapp H^{-1}$ and the above criteria are satisfied then the domains with a larger energy density disappear exponentially fast. This result is the basis of the devaluation mechanism.

\subsection{A numerical example}

As an illustrative example of the effect of a bias on the dynamics of the domain walls, we have solved Eqns. (\ref{13}--\ref{12}) numerically in the radiation era for a spherical domain wall, initially at rest, with a larger vacuum energy density in its inner domain. This allows us to determine the evolution of the domain wall's physical radius, $a q=R\gamma^{-1/2}$, until collapse (which was defined as the moment when $q$ vanishes). In this section we shall assume that $a_i=a(t_i)=1$ where $t_i=H_i^{-1}/2$ is the initial time (we take $t_i=1$). 

\begin{figure}
\includegraphics[width=3in]{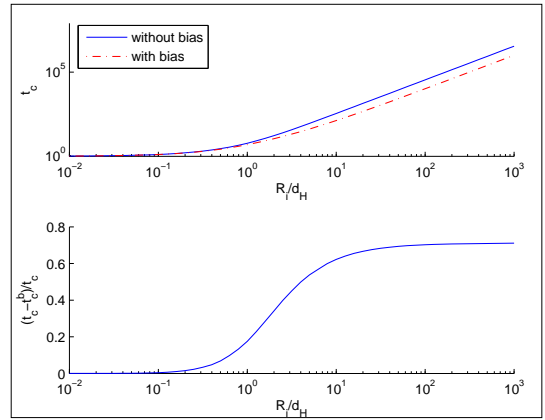}
\caption{\label{tempocolapso} The top panel shows the collapse time of a spherical domain wall, with and without an energy difference between the inner and outer domains, as function of the ratio between the initial wall invariant radius and the initial horizon. Time is in units of initial time, $t_i$. The lower panel depicts the relative difference between the time of collapse with and without bias. In both cases we have assumed $\epsilon/\sigma=1$.}
\end{figure}

Fig. \ref{tempocolapso} shows the time of collapse of a spherical domain wall and the relative difference between these collapse times as a function of the ratio between the initial wall invariant radius, $R$, and the initial Hubble radius. The bias term acts as a further mechanism which accelerates the domain walls, allowing them to overcome the Hubble damping term faster, and hence making them collapse in a shorter period of time. The relative importance of this effect grows as the initial radius increases. However, the curvature term in the equations becomes negligible for large $R$ and therefore in this limit the collapse of the wall is essentially determined by the bias term. As a result, in this limit the relative difference tends to a constant.

\begin{figure}
\includegraphics[width=3in]{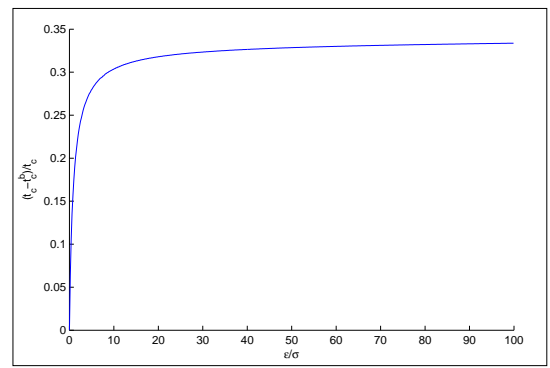}
\caption{\label{biasvariavel} The relative difference between time of collapse with and without bias, for a spherical domain wall with $R_i=H_i^{-1}$, for ${\epsilon}/{\sigma}$ ranging from 0 to 100.}
\end{figure}

Fig. \ref{biasvariavel} shows the relative difference between the collapse time, with and without an energy difference between minima, for ${\epsilon}/{\sigma}$ ranging from $0$ to $100$ (for a spherical domain wall with $R_i=H_i^{-1}$). Notice that as the bias term's importance grows, this relative difference tends to a constant. Since the bias term makes the velocity ultrarelativistic faster, the time of collapse with bias has an obvious lower bound, given by the time a photon takes to travel a distance equal to the wall's physical radius. In a flat FRW universe: $t_c^{min} = (1+q_i/2)^2$. For this particular case, this yields $t_c^{min} \sim 4$, and the relative difference approaches $\Delta \sim 0.33$ as the ratio ${\epsilon}/{\sigma}$ increases.

\section{The devaluation mechanism}

Devaluation \cite{Freese} is a mechanism which aims to explain the observed small non-zero value of the cosmological constant using the dynamics of biased domain walls. The key physical idea is that under plausible circumstances a network of unstable domain walls might form at a certain critical temperature, dividing the universe in many different regions with different values of the vacuum energy density. Then domain walls separating different vacuum domains will feel a pressure which will tend to suppress those with higher energy thus driving the universe towards lower and lower values of the vacuum energy density.

\subsection{A simple realization}

A simple potential that has the relevant features is
\begin{equation}
V(\phi)=V_0\cos{\left(\frac{\pi \phi}{\eta}\right)}-\epsilon\frac{\phi}{2\eta}+C\,,\label{toypotential}
\end{equation}
where $V_0$ is effectively the barrier height and $\epsilon$ is the energy difference between minima. From the original paper \cite{Freese}, the expectation is that
\begin{equation}
\eta\sim V_0^{1/4}> 10 \, {\rm MeV}\,\label{etabound}
\end{equation}
and 
\begin{equation}
\epsilon^{1/4}\sim\rho^{1/4}_{crit}\sim10^{-3}\,{\rm eV}\,.\label{epsilonbound}
\end{equation}
It is important to realize that, in this particular model, the domain walls interpolating between successive pairs of minima in the above potential all have similar tensions. The model also requires a large number of minima, but the exact number is actually not relevant for the analysis that follows. It was argued that domains with negative energy vacua would be suppressed so that domain wall dynamics would lower the cosmological constant to a value just below zero ($\rho_{vac} \sim \epsilon \sim \rho_{crit}$). We will question the validity of this assumption later on.

The initial conditions are expected to be such that $L_i\sim T_G^{-1}\sim V_0^{-1/4}\sim\eta^{-1}$, which is significantly smaller than the Hubble radius,
\begin{equation}
\frac{d_H}{L_i}\sim\frac{m_{Pl}}{V_0^{1/4}}\lapp10^{20}
\end{equation}
Clearly, with $L_i\sim\eta^{-1}$, the surface pressure $p_T\sim\eta^4\sim V_0$ will initially dominate the volume pressure $p_V\sim\epsilon$ and the walls will initially be very stable. So for this choice of parameters, devaluation is not very efficient. 

On the other hand, the largest correlation length we can have (which corresponds to the largest instability) is $L\sim t$, and recalling that
\begin{equation}
t\sim\frac{m_{Pl}}{T^2}
\end{equation}
in the radiation era and
\begin{equation}
t\sim\frac{m_{Pl}}{T^{3/2}T_{eq}^{1/2}}
\end{equation}
in the matter era, we find that in this maximally instability case the 
domain walls become unstable and disappear will for
\begin{equation}
\left(\frac{T}{m_{Pl}}\right)^2\sim\frac{\epsilon}{V_0}\frac{\eta}{m_{Pl}}
\end{equation}
\begin{equation}
\left(\frac{T}{m_{Pl}}\right)^{3/2}\sim\frac{\epsilon}{V_0}\frac{\eta}{m_{Pl}}\left(\frac{m_{Pl}}{T_{eq}}\right)^{1/2}\,.
\end{equation}
For the decay to happen in the radiation era we need $T>T_{eq}\sim 1 \, {\rm eV}$, and using Eq. (\ref{epsilonbound}) we find
\begin{equation}
\eta_{rad}\lapp 100 \, {\rm keV}\,,\label{radiationbound}
\end{equation}
which is clearly incompatible with the assumed bound given by Eq. (\ref{etabound}). It follows that if this mechanism (at least as originally presented) is to explain the observed value of the cosmological constant, the domain walls must survive through the radiation era, and decay only in the matter era. \textit{In other words, in the original scenario devaluation must occur late in the history of the universe, and not early.} This is due to the energy scale required to match the observed value of the cosmological constant, and is another manifestation of the underlying fine-tuning. Repeating the calculation for a decay during the matter era (and ignoring the effect of the recent acceleration phase on the expansion rate, which is negligible in this context) we now find
\begin{equation}
\eta_{mat}\lapp10 \, \rm MeV\,,\label{matterbound}
\end{equation}
which saturates the bound given by Eq. (\ref{etabound}).

Therefore the best we can do is to have a network that disappears around today. If so, and again neglecting the effect of the recent dark energy domination, we would expect the cosmological constant to be
\begin{equation}
\frac{\rho^{1/4}_{vac}}{m_{Pl}}\sim\left(\frac{\eta}{m_{Pl}}\right)^{3/4}\left(\frac{T_0}{m_{Pl}}\right)^{3/8} \left(\frac{T_{eq}}{m_{Pl}}\right)^{1/8}\,,
\end{equation}
and since $T_{eq}\sim 1 \, {\rm eV}$ and $T_0\sim2\times 10^{-4}\, {\rm eV}$ 
we get
\begin{equation}
\rho^{1/4}_{vac}\sim\left(\frac{\eta}{m_{Pl}}\right)^{3/4}10^{13} \, \rm eV\,,
\end{equation}
from which we would get $\rho_{vac}^{1/4}=10^{-3} \, \rm eV$ for $\eta\sim10 \, \rm MeV$ as previously stated.

There is, however, an obvious problem with such a scenario: $10 \, \rm MeV$ domain walls decaying today are observationally ruled out. For the rather classic Zel'dovich bound \cite{ZEL}, $\eta\sim 1 \, \rm MeV$, we only get $\rho_{vac}^{1/4}\sim 10^{-4}\, \rm eV$. This is optimistic both in the sense that the observational bound is somewhat lower, and that we really want the network to decay a bit before today to be clear of observational problems.

Indeed, for late devaluation there are several extra requirements (not present for early devaluation) which need to be satisfied. If devaluation is not complete and there are some walls still around, their average contribution to the energy density of the universe needs to be $\rho_{walls}\lapp 10^{-5} \rho_{crit}$. Otherwise, assuming that the characteristic size of the domains is $\gapp H^{-1}/100$, there would be a detectable contribution of domain walls to the cosmic microwave background anisotropies. This means that the wall tensions are strongly constrained. We know from Eqn. (\ref{decaybound}) that if we have two contiguous domains with a vacuum energy difference of $\epsilon$ then the domain with larger vacuum density will be exponentially suppressed when $\epsilon \gapp \sigma / L$. But $\sigma/L$ is the average energy density of the domain wall network which, as we saw, must be at least five orders of magnitude smaller than the critical density. So in this case, we obtain
\begin{equation}
\rho_{vac}^{1/4}\lapp 10^{-4} \, \rm eV\,,
\end{equation}
and devaluation would lead to a vacuum energy density significantly smaller than the critical density at the present time. In fact this number might even be smaller, depending on the domain wall tensions.

\subsection{General features of the mechanism}

Beyond particular realizations of the devaluation mechanism, several general comments can be made about the underlying physical scenario. First, let us note that this simple model may allow for positive and negative values of the vacuum energy density. It is argued \cite{Freese} that gravitation itself may prevent the vacuum energy density from attaining negative values. However, here we will be mainly concerned with the evolution of domain wall networks during the matter and radiation eras in which the contribution of the domain wall and the vacuum energy densities can be neglected. In this context, there is no cutoff that prevents the vacuum energy densities from attaining negative values. Such a mechanism can only operate when the vacuum energy densities are the dominant contribution for the dynamics of the universe. Hence, a low-energy cutoff of the order of the critical density at the present time needs to be introduced by hand in the devaluation model, which is clearly not an attractive feature of the model.

In this scenario a domain wall network forms at a critical temperature $T_c \sim V_0^{1/4}$. It is straightforward to put a lower bound the number, $N$, of domains that are initially present in a region with a comoving size of the order of the Hubble radius at the present day,
\begin{equation}
N \gapp \left(\frac{T_{\rm eq}}{T_0}\right)^{3/2} \left(\frac{T_{\rm c}}{T_{\rm eq}}\right)^3\,.
\end{equation}
Assuming a fixed energy difference between successive minima and that the barrier height and tuning of the potential are roughly comparable as in the simple devaluation toy model introduced above we see that the difference in energy densities between successive minima is bounded from below
\begin{equation}
\epsilon \lapp \frac{T_c^4}{N} \sim T_c (T_0 \, T_{\rm eq})^{3/2}
\end{equation}
if all the minima are populated at the time when the network forms. In fact, if we assume that close to $T_c$ the domains rapidly attain a typical size of the order of the Hubble radius then
\begin{equation}
\epsilon \sim T_c (T_0 \, T_{\rm eq})^{3/2}
\end{equation}
which is much larger that the energy density at the present time. Of course, not all the minima need to be populated and in this case $\epsilon$ can be smaller. However, in this case the domain wall tension is no longer the same for all domain walls since domain walls will in general interpolate between distant minima. We also note that a single domain with the lowest possible energy density does not necessarily survive domain wall evolution since, even in the absence of the devaluation mechanism (which will only operate for $\epsilon \gapp \sigma/L$), domain wall dynamics in the scaling regime naturally leads to the suppression of most of the available domains each Hubble time.

The above analysis also confirms the naive expectation that the devaluation mechanism can only be effective if $L \lapp H^{-1}$. If this is not the case (either because the walls are somehow pushed outside the horizon, or in the opposite limit, $L \ll H^{-1}$) the dynamics of the domain walls is even less efficient in suppressing domains with larger values of the vacuum density. In the absence of friction, and assuming that $L \lapp H^{-1}$, a domain wall network will in fact approach a scaling regime with $L \propto H^{-1}$, which is quite generically an attractor solution \cite{SIMS1,AWALL,IDEAL3}. Friction may slow down the domain walls, and if they are sufficiently light the evolution may be friction dominated up to the present time, but this will not help.

Therefore the devaluation scenario does not naturally lead to the required value of the vacuum energy density. A simple and physically intuitive way of expressing this is in terms of fine-tuning. A cosmological constant in the standard scenario is considered unappealing because the value that observations indicate for its magnitude is many orders of magnitude below what one would naturally expect from particle physics considerations. From this perspective the motivation of the devaluation scenario is that it would lead to an otherwise extremely small value in a natural way. However, this is not so because the same fine-tuning problem is still there.

Of course many other features of the simplest implementation of devaluation can be relaxed. We may have complex domain wall networks with junctions, domain walls with tensions that may be correlated with the differences between the vacuum energy densities. In fact the domain walls may not all be formed at the same time with additional domains with smaller energy densities separated by low tension domain walls being formed only at smaller critical temperatures. Still, it is clear that the devaluation mechanism is too efficient and consequently it will always be necessary to introduce by hand a low energy cut-off of the order of the critical density at the present day.

\section{\label{conc}Discussion}

In more realistic particle physics scenarios there will typically be many coupled scalar fields that can lead to domain walls. From a phenomenological point of view, there are several reasons why this more general case may differ from the simplest implementation of the devaluation mechanism. The potential may be significantly more complicated, as in landscape-type scenarios. This by itself need not be a great advantage, since in any case energy minimization criteria will always favor evolution down the potential, regardless of the number of fields. (In particular, any number of uncoupled fields will lead to a scenario very similar to the single-field case.) More important, though, is the existence of coupled fields, since this generically leads to domain wall networks with junctions, and also to more complicated spectra of wall tensions, which need not necessarily be of comparable magnitude. 

The presence of walls with significantly different tensions may be important for the network dynamics. Recall that biased walls will decay when their characteristic size grows to $L \gapp \sigma / \epsilon$. Since in this model $\epsilon$ is effectively fixed to the observed vacuum energy density, we see that, according to this criterium, higher-tension walls are actually more stable than lower-tension walls (since a larger characteristic size is needed to make them decay). Notice that this is contrary to standard energy minimization arguments, whereby higher-tension walls tend to decay into lower-tension walls. 

If we now consider a network of walls with junctions and a non-trivial hierarchy of tensions, and temporarily assume that the various types of walls with different tensions have comparable characteristic sizes (which may be too naive an assumption), then one may reach a threshold where the lowest-tension walls present become unstable and decay. This may render the whole network unstable and make it disappear well below what one would expect from a stability analysis for the higher-tension walls. Therefore this mechanism may increase the efficiency of devaluation. However, given that the devaluation mechanism is already too efficient in its simplest implementation this extra efficiency will not help. Hence a low energy cut-off of the order of the critical density at the present day is needed in order for devaluation to stop at the observed value of the dark energy density.

Although in this article we have dealt with biased domain walls in a post-inflationary context, it is important to realize that in the context of string landscape scenarios they can arise during the inflationary phase itself, as the inflationary field samples multiple non-degenerate vacua \cite{LANDSC}. From our point of view this would be an example of early devaluation, with the additional difference that the energy scales involved are much higher (and the walls would typically be horizon-sized). The subsequent evolution of the system will depend on several factors, including the relative populations of the different vacua and the scales of the various deviations from vacuum degeneracy. Requiring that the standard cosmological behavior is recovered by the time of nucleosynthesis leads to non-trivial constraints on the model parameters.

\begin{acknowledgments}
We thank Katherine Freese, Josinaldo Menezes, Roberto Menezes, Joana Oliveira 
and Douglas Spolyar for useful discussions. The work of C.M. is funded by a Ciencia2007 Research Contract. L. Sousa is supported by FCT through the grant SFRH/BD/41657/2007.
\end{acknowledgments}

\bibliography{biased}
\end{document}